\documentclass{aa}  
\usepackage{graphicx}
\usepackage{txfonts}
\usepackage{mathrsfs}
\usepackage{epic}  
\usepackage{eepic}
\usepackage{pstricks}
\usepackage{amssymb}
\usepackage{amsmath}
\usepackage{bm}
\usepackage{multirow}
\usepackage{textcomp}
\usepackage[]{natbib}
\usepackage{todonotes}
\usepackage{subfig}
\usepackage{booktabs}
\usepackage{placeins}
\usepackage{xcolor,colortbl}
\usepackage{pifont}
\newcommand{\cmark}{\ding{51}}
\newcommand{\xmark}{\ding{55}}
\newcommand{\bxi}{\pmb{ \xi}}
\newcommand{\AtmoModel}{\textit{Atmo}}
\colorlet{mygray}{gray!40}

\title{
   Atmospheric radiation boundary conditions for high frequency waves in time-distance helioseismology 
}
\titlerunning{
  Atmospheric radiation boundary conditions
}
\authorrunning{
  Fournier et al.
}
\author{
  D.~Fournier\inst{1}\thanks{Corresponding author: fournier@mps.mpg.de}, M.~Legu\`ebe\inst{1,2}, C.~S. Hanson\inst{1}, L.~Gizon\inst{1,2,3}, H.~Barucq\inst{4}, J.~Chabassier\inst{4},  M.~Durufl\'e\inst{5}
}
\institute{
  Max-Planck-Institut f\"ur Sonnensystemforschung, 
  Justus-von-Liebig-Weg 3, 37077 G\"ottingen, Germany
  \and 
  Institut f\"ur Astrophysik, Georg-August-Universit\"at G\"ottingen, 
  Friedrich-Hund-Platz 1, 37077 G\"ottingen, Germany
  \and
 Center for Space Science, NYUAD Institute, New York University Abu Dhabi, PO Box 129188, Abu Dhabi, UAE
  \and 
  Magique-3D, Inria Bordeaux Sud-Ouest, Universit\'e de Pau et des Pays de l'Adour, 64013 Pau, France
  \and 
  Magique-3D, Inria Bordeaux Sud-Ouest, Universit\'e de Bordeaux, 33400 Talence, France
}
\date{
  Received \today; accepted XXX
}

\abstract
  {
 The temporal covariance between seismic waves measured at two locations on the solar surface is the fundamental observable in time-distance helioseismology. Above the acoustic cut-off frequency ($\sim$5.3~mHz), waves are not trapped in the solar interior and the covariance function can be used to probe the upper atmosphere. 
   We wish to implement appropriate radiative  boundary conditions for computing the propagation of high-frequency waves in the solar atmosphere. 
We consider the radiative boundary conditions recently developed by Barucq et al. (2017) for atmospheres in which sound-speed is constant and density decreases exponentially with radius. We compute the cross-covariance function using a finite element method in spherical geometry and in the frequency domain.  The ratio between first- and second-skip amplitudes in the time-distance diagram is used as a diagnostic to compare boundary conditions and to compare with observations.
We find that a boundary condition applied 500 km above the photosphere and derived under the approximation of small angles of incidence accurately reproduces the `infinite atmosphere' solution for high-frequency waves.
When the radiative boundary condition is applied 2 Mm above the photosphere, we find that the choice of atmospheric model affects the time-distance diagram. In particular, the time-distance diagram exhibits double-ridge structure when using a VAL atmospheric model. 
}


\newcommand{\g}{\gamma}

\newcommand{\w}{\omega}

\newcommand{\e}{\textrm{e}}
\newcommand{\ii}{\textrm{i}}

\newcommand{\Ecal}{\mathcal{E}}

\newcommand{\RSUN}{R_\odot}
\newcommand{\RA}{r_\mathrm{t}}

\newcommand{\WAC}{\w_{\mathrm{c}}}
\newcommand{\oWAC}{\omega_{\mathrm{t}}}

\newcommand{\KH}{k_\mathrm{h}}

\newcommand{\BR}{\bm{r}}
\newcommand{\ZHAT}{\hat{\bm{z}}}

\newcommand{\LBO}{\Delta_\mathrm{h}} 

\newcommand{\abs}[1]{\left|#1\right|}

\newcommand{\ddp}[2]{\dfrac{\partial#1}{\partial#2}}

\newcommand{\referee}[1]{#1}
   
\begin{document} 

\maketitle

\section{Introduction}

The interpretation of local helioseismology observations 
requires a description of wave propagation in the Sun (the forward problem). Solar seismic waves are described by a wave operator and  boundary conditions. The boundary conditions are important, in particular if one wants to study waves with frequencies above the acoustic cutoff frequency.

Common practice in helioseismology is to use a free-surface boundary condition \citep[e. g.][]{jcd_etal_1996,bogdan_etal_1996}; but this approach forces all waves to reflect and thus are not appropriate for high frequency waves. 
One solution to model outgoing waves 
is to introduce Perfectly Matched Layers (PML), first introduced by \citet{BER94}. This method consists of adding an extra layer to the computational domain in which the waves are artificially damped by some user-defined factor. Applications have included the modeling of waves in stratified media in geophysics \citep{KOM07}, as well as solar MHD wave simulations \citep{Cameron2008, HAN10, SAN15}. Applications of PMLs in helioseismology have shown that, though the boundary layer absorbs the waves as expected, this is at the expense of two compromises. Firstly, the addition of a layer increases the computational burden significantly to capture the small wavelengths in the upper atmosphere. Second, the PMLs substitute a constant infinite medium which is not the case for the Sun.
The purpose of this paper is to propose and implement a radiative boundary condition (RBC) that is not subject to these two limitations.

We consider a scalar wave equation for the divergence of the wave displacement $\psi(\BR,\omega) = \rho(\BR) c^2(\BR) \nabla \cdot \bxi(\BR, \omega)$, which in the frequency domain is given by
\begin{equation}\label{eq:main}
  -\frac{\sigma^2}{\rho c^2}\psi 
  - \nabla\cdot\left(\dfrac{1}{\rho}\nabla \psi \right) = s,
\end{equation}
where 
\begin{equation}
\sigma^2 = \w^2+2\ii\w\gamma
\end{equation}
depends on mode frequency $\omega$ and attenuation $\gamma(\omega)$.
This equation, of Helmholtz type, was studied in detail by 
 \citet{Gizon2017} and is appropriate to study acoustic waves that propagate in the solar convection zone. The medium is specified by the sound-speed $c$, and the density $\rho$, which are read from the standard solar Model~S \citep{jcd_etal_1996}.
The source function $s(\BR, \omega)$ excites the waves. The development of RBCs for the Helmholtz equation began with \citet{BAY82}, but was designed under the assumption that a homogeneous medium is situated outside of the computational domain, which is not the case for the Sun. In order to overcome this difficulty, \citet{Gizon2017} extended and smoothed Model~S up to 4~Mm above the photosphere in order to reach a constant density and sound speed, where a Sommerfeld boundary condition was applied. While this method has produced a solar-like power spectrum, it requires a significant increase in computational cost in order to model the additional layer in the atmosphere. Recently \citet{Barucq2017} proposed a set of radiative boundary conditions for 
an atmosphere in which the sound speed is constant and the density decays exponentially. In this paper, we study these proposed boundary conditions in the context of time-distance helioseismology.

In Sect.~\ref{sec:eq}, we describe the equations that we are solving for 1D and 2D background models. In Sect.~\ref{sec:BCs} we describe the radiative boundary conditions from \citet{Barucq2017}. Section~\ref{sec:numericalImplement} details the numerical method used to solve Eq.~\eqref{eq:main}, and section~\ref{sec:results} compares the various boundary conditions properties and their absorption properties.  In Sect.~\ref{sec:resultsHF} we extend the atmosphere with a VAL atmospheric model and study its influence on the time-distance diagram at high frequencies.

\section{Scalar wave equations}
\label{sec:eq}

  \subsection{2D background models}

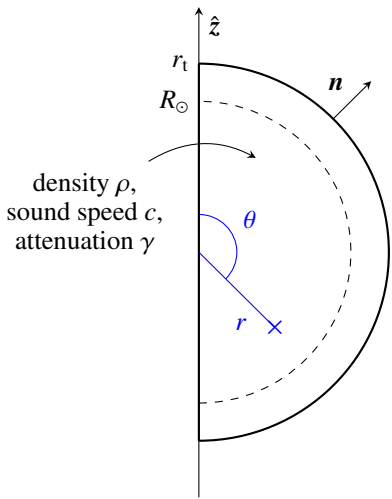
\begin{figure}[h]
\centering
\begin{tikzpicture}[x=2.5cm,y=2.5cm]
  \draw [-stealth](0,-1.3) -- (0,1.3) node[below right]{$\ZHAT$};
  \draw[thick] (0,-1.) arc (-90:90:1) node[left]{$\RA$};
  \draw[thick] (0,-1) -- (0,1);
  \draw[dashed] (0,-0.8) arc (-90:90:0.8) node[left]{$\RSUN$};
  \node[blue] at (0.4,-0.4) {$\times$};
  \draw[blue] (0.4,-0.4)-- (0,0) node [pos=0.25,below left]{$r$};
  \draw[blue] (0,0.2) arc (90:-45:0.2) node [pos=0.5,above right]{$\theta$};
  \node (P) at (-0.6,0.2) {\begin{tabular}{c}density $\rho$,\\ sound speed $c$,\\  attenuation $\gamma$\end{tabular}};
  \draw[-stealth] (node cs:name=P,angle=40) to[bend left] (0.3,0.5);
  \draw[-stealth] (0.7071,0.7071) -- + (0.2,0.2) node[pos=0.5,above left]{$\bm{n}$};
\end{tikzpicture}
\caption{Geometrical setup for 2.5D computations. The computational domain is delimited by thick lines and the photosphere by the dashed line.}
\label{fig:geom}
\end{figure}

Without any assumption, the domain in which Eq.~\eqref{eq:main} is defined is three dimensional and encompasses the entire Sun. It is however reasonable to assume that the solar background properties are symmetric about the rotation axis ($\ZHAT$). Under this assumption Eq.~\eqref{eq:main} can then be projected onto azimuthal modes 
\begin{equation}
\psi(\BR) = \sum_m \psi_m(r,\theta) \e^{\ii m\phi} .
\end{equation}
Each mode $\psi_m$ is then the solution of a 2D partial differential equation,
\begin{equation}
-\frac{\sigma_m^2}{\rho c^2} \psi_m  - \tilde{\nabla}\cdot\left(\dfrac{1}{\rho}\tilde{\nabla} \psi_m \right) = s_m(r,\theta), \label{eq:main2D}
\end{equation}
where $\sigma_m^2 = \w^2 + 2\ii\w\g - m^2 c^2 / (r\sin\theta)^2$ and $\tilde{\nabla} = \hat{\BR} \partial_r + \hat{\boldsymbol \theta} \frac{1}{r} \partial_\theta$ is the 2D spatial gradient operator and $\tilde{\nabla} \cdot \mathbf{f} = \partial_r (r^2 f_r) /r^2 +  \partial_\theta(\sin\theta f_\theta) / r\sin\theta$ is the 2D divergence operator. The source term $s_m$ represents the projection of $s$ onto the azimuthal modes. Figure~\ref{fig:geom} shows the resultant geometrical setup. Here, $R_\odot$ denotes the solar radius and $\RA$ the limit of the computational domain where an artificial boundary condition needs to be applied.
    
\subsection{1D background models}

When the background model depends only on radius, the solution can be written in terms of spherical harmonic functions:
\begin{equation}
\psi(\BR) = \sum_{\ell=0}^\infty \sum_{m=-\ell}^\ell \psi_{\ell m}(r) Y_{\ell}^m(\theta, \phi).
\end{equation}
For each $\ell$ and $m$ the function $\psi_{\ell m}(r)$ is defined on the 1D domain $0\le r \le \RA$ and solves
\begin{equation}
- \left( \frac{\sigma^2}{c^2} - \KH^2 \right)\psi_{\ell m} - \frac{\rho}{r^2} \frac{d}{dr} \left( \frac{r^2}{\rho}  \frac{d \psi_{\ell m}}{dr} \right) = \rho s_{\ell m}(r), \label{eq:main1D}
\end{equation}
where 
\begin{equation}
\KH = \frac{\sqrt{\ell(\ell+1)}}{r}
\end{equation}
is the local horizontal wavenumber, and $s_{\ell m}$ is the projection of the source $s$ onto spherical harmonics.

\subsection{Atmospheric model}

\begin{figure}
\includegraphics[width=\linewidth]{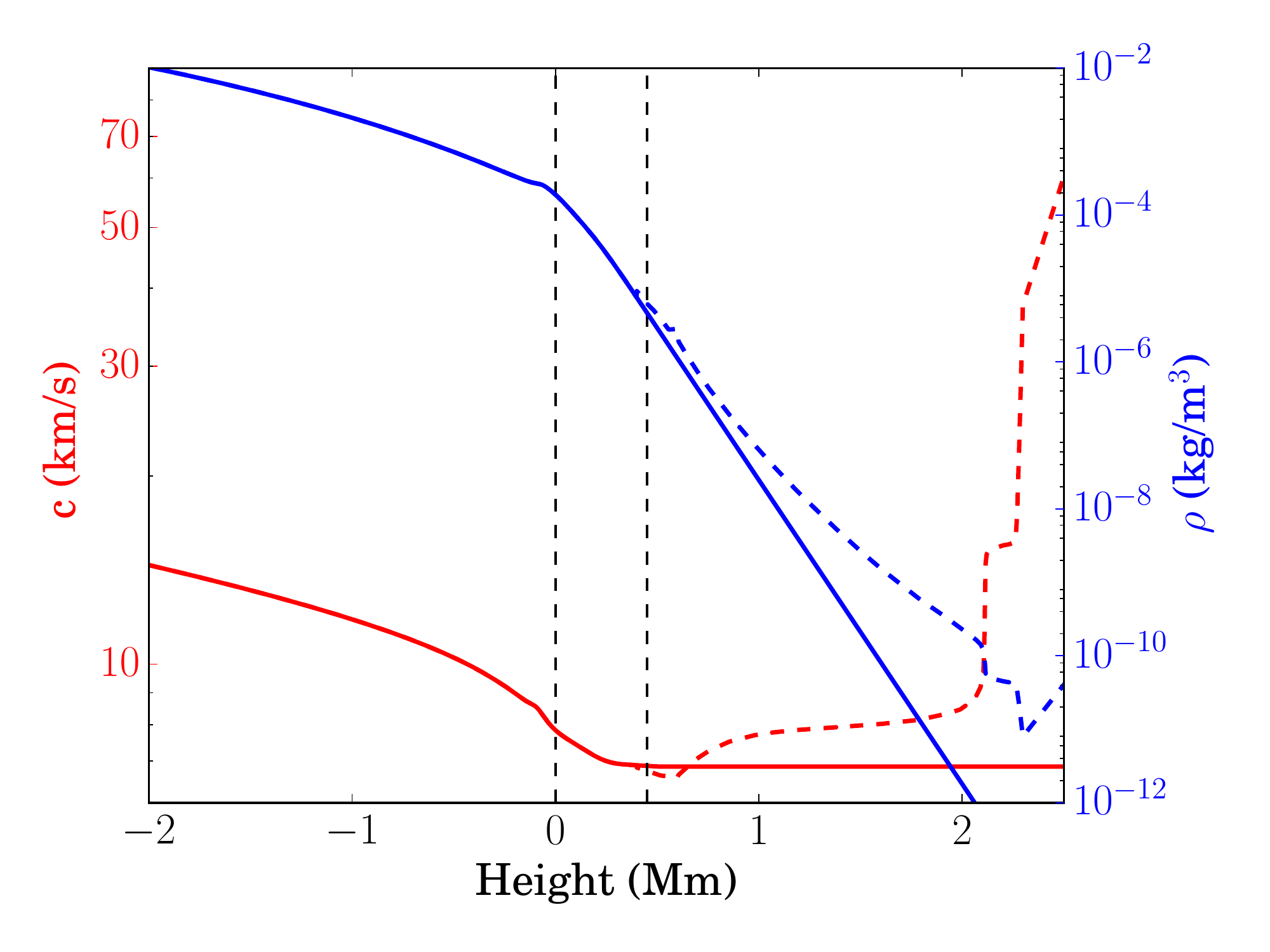}
\caption{Sound speed (red) and density (blue) in the upper layers of the Sun as functions of height measured from the photosphere. Below 500~km (vertical dashed line), the sound speed and density are from Model~S \citep{jcd_etal_1996}. Above 500~km, the model \AtmoModel \ (solid lines) consists of an extension where density decays exponentially ($H = 105$~km) and  sound speed is constant ($c = 6855$~m/s). The dashed lines show the VAL-C atmospheric model up to a height of $2500$~km \citep{VER81}, which will be used in Sect.~\ref{sec:resultsHF}.} \label{fig:rhoc}
\end{figure}

Here we discuss the background atmosphere used in this study. The sound speed and density are taken from the standard Model~S \citep{jcd_etal_1996} which extends up to 500~km above the solar surface. We extend this model by supposing that the density continues to decay exponentially at the same rate as at the end of Model~S (density scale height equals 125~km). Additionally, the sound speed is smoothed to a constant value of 6855~m/s. We will refer to this model as \AtmoModel. This model is similar to the one presented by \citet{Schunker2011} but with a  different density scale height (125~km compared to 105~km). While \AtmoModel~is reasonable to model the lower chromosphere, it does not reproduce the temperature jump at the transition region around 2~Mm proposed in atmospheric models such as the VAL models \citep{VER81}. Plots of the sound speed and density for \AtmoModel \ and VAL-C are given in Fig.~\ref{fig:rhoc}. A smooth transition between the model~S and the VAL-C atmosphere is done between 400 and 600~km. 

\section{Radiative boundary conditions}
\label{sec:BCs}
\subsection{Radiative boundary condition for 1D background}

It is possible to transform Eq.~\eqref{eq:main1D} in order to obtain an Helmholtz equation that is more convenient to explain the behavior of high frequency waves. This can be done by changing the unknown,
\begin{equation}
\phi_{lm} = r \rho^{-1/2} \psi_{lm} \label{eq:phi},
\end{equation}
 resulting in a new form for Eq.~\eqref{eq:main1D},
 \begin{equation}\label{eq:main1Dv2}
 \frac{d^2\phi_{lm}}{dr^2} + Q \phi_{lm} = -r\rho^{1/2} s_{lm},
 \end{equation}
where 
\begin{equation}
Q = \frac{\sigma^2-\WAC^2}{c^2} - k_h^2 \label{eq:K2}
\end{equation}
and $\WAC$ is a cut-off frequency defined by
\begin{equation} \label{eq:Wac}
 \frac{\WAC^2}{c^2} =  \dfrac{1}{4H^2} \left(1 - 2 \frac{dH}{dr} + \frac{4H}{r} \right),
\end{equation}
where
\begin{equation}
H(r) = - \left( \dfrac{\mathrm{d}\,\ln\rho}{\mathrm{d}r} \right)^{-1}
\end{equation}
is the density scale height.

\begin{figure}[h]
\centering
\includegraphics[width=\linewidth]{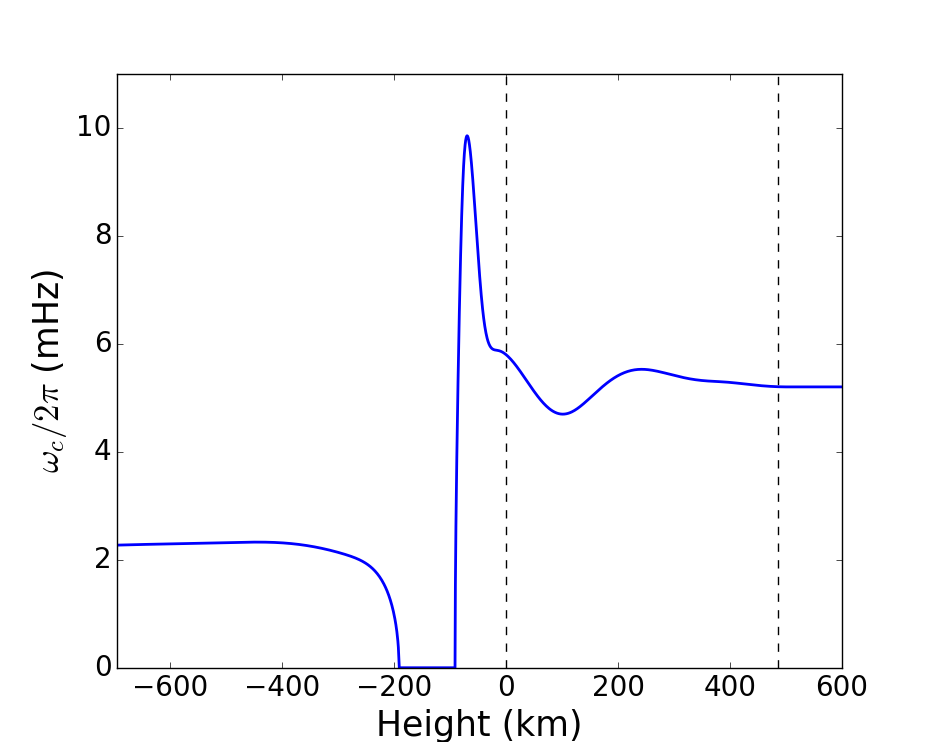}
\caption{
Acoustic cut-off frequency $\WAC$ as a function of height from the photosphere. Below 500~km, $\WAC$ is obtained using Model~S and Eq.\eqref{eq:Wac}; the density scale height is obtained after smoothing the density. Above 500~km, the acoustic cut-off frequency is constant and equal to $\oWAC = 5.3$~mHz. $\WAC$ is complex near $~-200$~km, although $\WAC^2$ is always positive and only $\WAC^2$ enters the wave equation.}
\label{fig:wac}
\end{figure}

A representation of $\WAC$ is given in Fig.~\ref{fig:wac}. With \AtmoModel, the density scale height is constant in the new region, $H = 125$~km, resulting in a constant acoustic cutoff $\oWAC / 2\pi = 5.3$~mHz. 
The wave behavior is determined by the sign of the real part of $Q$. The low frequency waves ($\omega < \oWAC$) are reflected back in the Sun, while the  high frequency ones ($\omega > \oWAC$) tunnel through the peak at 100~km below the surface and continue propagating into the upper atmosphere. 

With Eq.~\eqref{eq:main1Dv2} in hand, a radiative boundary condition is given by
\begin{equation}
\frac{d \phi_{lm}}{dr} = \ii Q^{1/2} \phi_{lm} \; \text{at } r = \RA. \label{eq:BCphi}
\end{equation}
Note that $Q$ can be negative and thus $Q^{1/2}$ is imaginary for frequencies above the acoustic cut-off $\oWAC$.
Using Eqs.~\eqref{eq:phi} and \eqref{eq:BCphi}, we can write the radiation boundary condition for the initial unknown $\psi_{lm}$
\begin{equation}\label{eq:AtmoNL}\tag{Atmo Non Local}
 \dfrac{d\psi_{\ell m}}{dr} = \left[  \ii Q^{1/2} -\left(\dfrac{1}{\RA}+\dfrac{1}{2H}\right) 
 \right] \psi_{\ell m} \; \text{at } r = \RA.
\end{equation}
This boundary condition (referred herein as \ref{eq:AtmoNL}) depends upon both the background model (density, sound speed) and the mode $\ell$ and its implementation is possible for any numerical method used to solve Eq.~\eqref{eq:main1D}. In the case of a homogeneous medium, it reduces to the Sommerfeld Radiation Condition where the derivative of the solution is linked to the solution itself via a wavenumber that depends on the properties of the medium.

\subsection{Approximate boundary conditions for the 2D problem.}
\label{sec:BC2D}

The implementation of the \eqref{eq:AtmoNL} boundary condition is not possible for 2D computations as the term $\ell(\ell+1)/\RA^2$ present in $Q$ should be replaced by the horizontal Laplacian $\LBO$
\begin{equation}
\LBO \psi_m = \frac{1}{r^2\sin\theta} \frac{\partial}{\partial \theta} \left( \sin\theta \frac{\partial \psi_m}{\partial \theta} \right) - \frac{m^2}{r^2 \sin^2\theta} \psi_m. 
\end{equation}
However, since this term is located under the square root, it leads to a spatially non-local operator which cannot be implemented directly. In order to overcome this difficulty an asymptotic expansion of the square-root is required. This can be achieved in one of two ways as detailed in \cite{Barucq2017}.

The first approach is to take a high-frequency approximation by making a first-order approximation of the square root supposing that $\oWAC/\omega$ is small,
\begin{equation}\label{eq:AtmoHF1}\tag{Atmo HF 1}
  \ddp{\psi_{m}}{r} = \left[ -\left(\dfrac{1}{\RA}+\dfrac{1}{2H}\right) + \dfrac{\ii\sigma}{c}\left(1-\dfrac{1}{2}\dfrac{\oWAC^2}{\sigma^2}\right) \right] \psi_{m}
 -\dfrac{\ii c}{2\sigma}\LBO \psi_{m}.
\end{equation}
Here, the horizontal Laplacian is not under a square root and thus can be implemented, for example, using the Finite Element Method (see next section).

The other approach is to expand around the angle of incidence with respect to the normal of the spherical surface. If we consider waves with $\KH \ll \omega / c$, the expansion of the square root leads to:
\begin{align}
  \ddp{\psi_m}{r} =& \left[ -\left(\dfrac{1}{\RA}+\dfrac{1}{2H}\right)  + \ii \left(\dfrac{\sigma^2-\oWAC^2}{c^2}\right)^{1/2} \right] \psi_m \notag\\ &\ - \dfrac{\ii}{2}\left(\dfrac{\sigma^2-\oWAC^2}{c^2}\right)^{-1/2} \LBO \psi_m.\label{eq:AtmoHAI1}\tag{Atmo SAI 1}
\end{align}
The general form of this BC is similar to \eqref{eq:AtmoHF1} since the radial derivative at the boundary $\ddp{\psi_m}{r}(\RA)$ is proportional to the sum of two terms: one linked to the solution itself $\psi_m$ and the other to the horizontal Laplacian $\LBO \psi_m$. The only difference between \eqref{eq:AtmoHF1} and \eqref{eq:AtmoHAI1} lies in the factors in front of these two terms. \referee{While the general form of these two boundary conditions is similar, we expect them to behave differently depending on the validity of the hypothesis made in their respective derivations.}

In the Finite Element Method, $\LBO \psi_m$ is straight forward to compute since it naturally arises from the weak-form (see next section). However, other methods may encounter difficulties in computing $\LBO \psi_m$ and thus we seek a further expansion of \eqref{eq:AtmoNL} to ease computational difficulties. A 0-th order expansion of \eqref{eq:AtmoNL} can be done by simply neglecting the term involving $\KH^2$ while not expanding around $1/\sigma^2$ as it was done for \eqref{eq:AtmoHF1}. This leads to the last BC we consider in this study:
\begin{equation}\label{eq:AtmoHAI0}\tag{Atmo SAI 0}
  \ddp{\psi_m}{r} = \left[ -\left(\dfrac{1}{\RA}+\dfrac{1}{2H}\right) + \ii \left(\dfrac{\sigma^2-\oWAC^2}{c^2}\right)^{1/2} \right] \psi_m.
\end{equation}
This condition is again a Sommerfeld-like boundary condition with a complex wavenumber that depends on the properties of the medium. However, unlike \eqref{eq:AtmoNL}, it does not take into account the horizontal wavenumber $\KH$. It is interesting to note that this boundary condition is similar to the one obtained by \citet{GOU93} by using a plane-parallel isothermal atmosphere
\begin{equation}
 \ddp{\psi_m}{r}(\RA) = \left[ \dfrac{1}{2H} + \ii \left(\dfrac{\sigma^2-{\WAC}_0^2}{c^2}\right)^{1/2} \right] \psi_m.
\end{equation}
The term in $1/\RA$ is missing due to the plane-parallel approximation and the acoustic cut-off ${\WAC}_0 = c / 2H$ is the isothermal one contrary to the one in \eqref{eq:AtmoHAI0}. 

A summary of the different boundary conditions is given in Table~\ref{tab:BCs}. It also includes the (Dirichlet) boundary condition which supposes that the Lagrangian perturbation of the pressure is zero at the surface and is often used in helioseismology
\begin{equation} \label{eq:Dirichlet}\tag{Dirichlet}
\psi_m(\RA) = 0.
\end{equation}
This BC is the limit of our model when the density in the atmosphere decays infinitely fast ($H \rightarrow \infty$). To test the boundary conditions presented thus far, we will use an `infinite' atmosphere where a Dirichlet boundary condition is imposed at $35$~Mm above the photosphere
\begin{equation} \label{eq:Reference}\tag{Reference}
\psi_m(\RA) = 0 \quad \textrm{at } r_t = \RSUN + 35 \textrm{ Mm}.
\end{equation}
In this way, all the waves traveling in the extended atmosphere never reach the boundary at 35~Mm due to the damping. This boundary condition is denoted as (Reference) for the remainder of this study.

All the proposed boundary conditions can be implemented for a 1D solver and present the same degree of complexity so one can use \eqref{eq:AtmoNL} as it represents exactly our atmospheric model. For a 2D problem, this exact boundary condition is no longer available and one should choose between a high-frequency or small angle of incidence approximation.

\begin{table}
\caption{Summary of the different boundary conditions used in this paper with the associated domain of validity and their applicability for 1D and 2D backgrounds.}
\begin{tabular}{cccc}
\hline
Boundary condition & Domain of validity & 1D & 2D \\
\hline
 (Reference)           & All cases & \cmark & \cmark \\
 \hline 
  \eqref{eq:AtmoNL}     &  All cases & \cmark & \xmark \\
  \eqref{eq:AtmoHF1}    & High frequency  & \cmark & \cmark \\
   \eqref{eq:AtmoHAI0}   & Small angle of incidence  & \cmark & \cmark \\
  \eqref{eq:AtmoHAI1}   & Small angle of incidence  & \cmark & \cmark \\
  \hline
 (Dirichlet)             & Free surface  & \cmark & \cmark \\ 
  \hline
\end{tabular}
\label{tab:BCs}
\end{table}

\section{Numerical implementation}
\label{sec:numericalImplement}

\subsection{Weak formulation of the equations}

In order to solve Eq.~\eqref{eq:main2D} for an azimuthally symmetric background or Eq.~\eqref{eq:main1D} for a spherically symmetric background, we use the Finite Elements Method (FEM). The implementation of the finite element method requires to write the weak form of Eqs.~\eqref{eq:main2D} or \eqref{eq:main1D}. In order to show how the boundary condition is implemented, we write the weak formulation for the 2D case (the derivation is similar in 1D). Let us multiply Eq.~\eqref{eq:main2D} by a test function $\phi$ and integrate over the whole domain
\begin{equation}
-\int_\Sigma \frac{\sigma_m^2}{\rho c^2} \psi_m \phi dS - \int_\Sigma \tilde{\nabla} \cdot \left( \frac{1}{\rho} \tilde{\nabla} \psi_m \right) \phi dS = \int_\Sigma s_m \phi dS.
\end{equation}
The term containing the Laplacian is then integrated by parts
\begin{equation}
\int_\Sigma \tilde{\nabla} \cdot \left( \frac{1}{\rho} \tilde{\nabla} \psi_m \right) \phi dS = - \int_\Sigma \left( \frac{1}{\rho} \tilde{\nabla} \psi_m \right) \cdot \tilde{\nabla} \phi dS + \int_{\partial \Sigma} \frac{1}{\rho} \partial_n \psi_m \phi dl,
\end{equation}
where the last term is a line integral over half a circle at the computational boundary (see Fig.~\ref{fig:geom}). The normal derivative $\partial_n \psi_m$ is then replaced by the appropriate boundary condition. If the boundary condition contains a horizontal Laplacian as \eqref{eq:AtmoHF1} or \eqref{eq:AtmoHAI1} then the surface term is further integrated by parts. 

For 1D and 2D backgrounds, the domain is decomposed into cells (segments in 1D, quadrangles in 2D) and the solution is determined as a polynomial on each cell. The degree of this polynomial determines the order of discretization of the method. The solution $\psi_m$ and the test function $\phi$ are expressed as a sum of polynomials on each cell which leads to the matrix form that is solved numerically (see e.g. the textbook by \citet{FEMbook}). For both types of background, we use the Montjoie code, developed at Inria\footnote{\texttt{http://montjoie.gforge.inria.fr}} and adapted to helioseismology by \citet{Gizon2017} to solve the scalar wave equation with solar parameters.

\subsection{Computational cost}

In order to demonstrate the efficiency gained by these new boundary conditions we compare the computational cost when the boundary of the domain $\RA$ is located at a height of 500~km (end of model~S) and at 4~Mm as in \citet{Gizon2017}. The mesh size is chosen in order to have at least 10 degrees of freedom per wavelength (1 cell with 10$^\mathrm{th}$ order polynomials). 

For the boundary condition at 500~km, we need $N_r = 33$ mesh points in the radial direction. For the 2D code, the mesh is built so that the cells have similar dimensions in the horizontal and vertical directions. This results in many cells in the $\theta$ direction close to the surface (see Fig.~6 in \citet{Gizon2017}) producing a 2D mesh consisting of approximately $10,000$ cells (see Table~\ref{tab:compCost}).

Extending the computational boundary to 4~Mm results in five additional layers of cells in the radial direction ($N_r = 38$). These cells are of equal height since the sound speed is constant. This extension adds about 50\% more cells to the mesh and results in an increase of the total used memory and computational time of 50\% as shown in Table~\ref{tab:compCost}. Thus, being able to simulate the propagation of the waves in the extended atmosphere and keeping the computational boundary at 500~km above the surface will result in a significant saving in computational burden.

\begin{table}
  \centering
  \caption{Number of mesh cells and corresponding computations times for solving the scalar wave equation Eq.~\eqref{eq:main2D} \referee{with or without extending the computational domain above Model~S.}
  }
  \label{tab:compCost}
  \small
  \begin{tabular}{ccccc}
    \midrule
    Outer radius $\RA$ & $N_r$ & Mesh cells & Memory & Comp. Time \\\midrule
    $\RSUN+4$~Mm   & 33 & 10749 & 6.6~GB & 43~s \\\midrule
    $\RSUN+500$~km & 38 & 6958 & 4.4~GB & 28~s \\\midrule
  \end{tabular}
  
\normalsize
\end{table}

\section{Comparison of the different boundary conditions}
\label{sec:results}

To compare the different boundary conditions, we compute the expectation value of the cross-covariance function. \citet{Gizon2017} showed that if the sources are spatially uncorrelated and equipartitioned then the expectation value of the cross-covariance is related to the imaginary part of the Green's function
\begin{equation} \label{eq:ImG}
\overline{C}(\BR,\BR_s,\w) = f(\w) \ \textrm{Im} G(\BR,\BR_s,\w).
\end{equation}
The function $f(\omega)$ controls the frequency dependence of the source covariance and will be used to filter waves in a given frequency range. Since waves of frequency below or above the acoustic cut-off frequency behave differently at the surface (where $\oWAC/2\pi \sim 5.3$~mHz), we use the function $f(\omega)$ to select waves centered around two frequencies, 
\begin{equation}\label{eq:PS}
  f(\w) = \exp\left(-\frac{(\abs{\w}-\w_0)^2}{2s^2}\right),
\end{equation}
where $\w_0/2\pi = 3$~mHz or $6.5$~mHz and $s/2\pi = 0.3$~mHz.

The Green's function $G$ is obtained by solving Eq.~\eqref{eq:main} with a Dirac delta function as source term located at $\BR_s$. 
For each boundary condition, Green's functions are computed for 7,200 frequencies uniformly distributed from 0 to 8.3~mHz, corresponding to a 10~day observational period at 60~s cadence. The waves are damped as a function of frequency according to the power law model of \citet{Gizon2017}.

To quantify the reflexivity of the different boundary conditions, we
fit two wavelets to the cross-covariance, one for each skip,
\begin{equation}
  C_{\textrm{fit}}(t) = \displaystyle\sum_{i=1,2}^{} A_i\exp\left(-\dfrac{(t-t^0_i)^2}{2\sigma_i^2}\right)\cos(\w_i t+\phi_i).
\end{equation}
We measure the energy of the outgoing (first skip) and reflected (second skip) wave packets as follows:
\begin{equation}\label{eq:energyDef}
  \Ecal_i(\BR, \BR_s) = \int_{w_i} \overline{C}(\BR,\BR_s, t)^2 \,\mathrm{d}t,
\end{equation}
where the intervals $w_i = [t_i^0 - 4\sigma_i, t_i^0+4\sigma_i]$ select the first- and second-skip wave packets.

\subsection{Exact atmospheric radiation boundary conditions}

As an initial test, the Dirac source is located at the center of the Sun. In this case, $\KH^2 = 0$ and the RBCs ~\eqref{eq:AtmoHAI1} and ~\eqref{eq:AtmoHAI0} are equivalent to~\eqref{eq:AtmoNL} and thus all boundary conditions should behave similarly.

\begin{figure*}[!htb]
\centering
\includegraphics[width=0.45\linewidth]{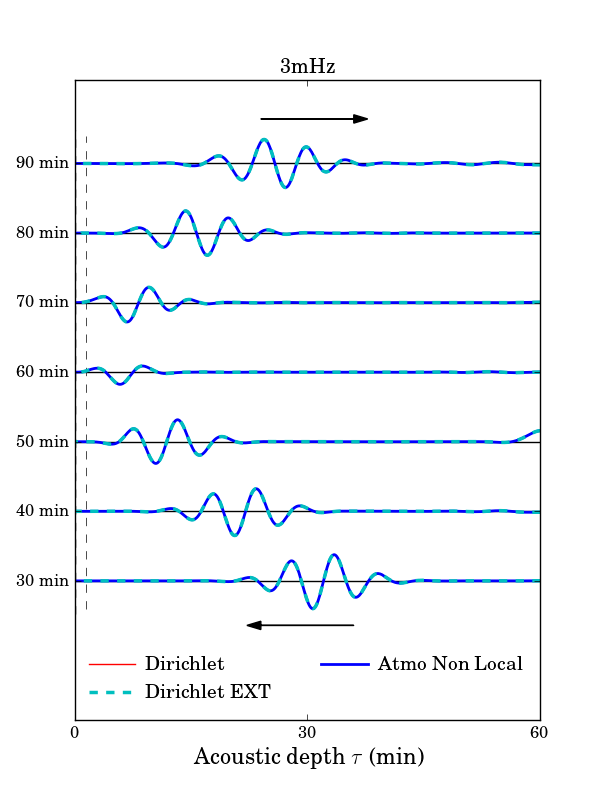}\quad
\includegraphics[width=0.45\linewidth]{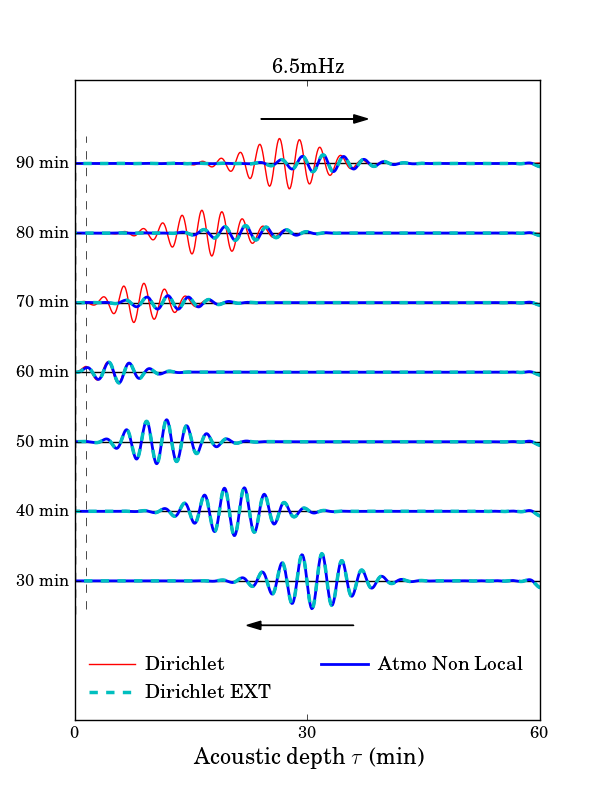}
\caption{
Temporal cross-covariance $c(r)^2 \overline{C}(0,r,t) \exp(t/90\text{ min})$ computed with~\eqref{eq:AtmoNL} and (Dirichlet) boundary conditions, originating from the center, with frequencies selected around 3~mHz (left panel) and 6.5~mHz (right panel). As a comparison, we also plot the cross-covariance when using the (Reference) BC (dashed cyan line). The signal is multiplied by $\exp(t/90\text{ min})$ to counteract the effect of the damping, and normalized by the maximum of its amplitude at $t=30$ min. The amplitude is plotted as a function of the acoustic depth $\tau(r) = \int_r^{r_a} \mathrm{d}r'/c(r')$ for better readability of near-surface behavior. The dashed line indicates the solar surface.}
\label{fig:Cxt}
\end{figure*}

Figure~\ref{fig:Cxt} shows the spatial evolution of the cross-covariance for wave packets centered at $\w_0/2\pi = 3$~mHz and $\w_0/2\pi = 6.5$~mHz. The wave packet propagates from the right to the left of the figure before being reflected and traveling in the other direction. It is shown that for any boundary condition, the low frequency signal is identically reflected near the solar surface.
The energy ratios between the outgoing and reflected wave packets are given in Table~\ref{tab:EnergyRatioCenter} for each boundary condition. For low frequency waves approximately 22\% of the energy of the outgoing packet is reflected (with energy loss due to damping) corresponding to a reduction in amplitude of about $47\%$. The lack in difference between each boundary condition comes at no surprise since the low frequency waves are reflected by the density profile before reaching the computational boundary. 

At frequencies above $\oWAC$, Dirichlet reflects the same quantity of energy as the 3~mHz case, though the cause for reflection, at this frequency, is the boundary and not the density profile. For the RBCs and (Reference) BC only 2\% comes back with the extended atmosphere. This weak reflection is not due to the RBCs but to the near-surface stratification profile and the fact that the packets are measured below the surface. The solutions using  the \eqref{eq:AtmoHAI1} and (Reference) BCs are identical, suggesting that the BC appropriately models the exponential decay of the density in the extended atmosphere as intended.  Finally, we note that the high frequency approximation made with the \eqref{eq:AtmoHF1} BC induces a small error (0.6\% difference) in the reflection coefficient. 

\begin{table}
  \small
  \centering
  \caption{
  Ratios between outgoing  and reflected energies ($\Ecal_2 / \Ecal_1$) and amplitudes ($A_2 / A_1$) for a source at the center of the Sun $r_s~=~0.$ and a receiver at $r=0.8~\RSUN$, for selected ranges of frequencies around 3~mHz and 6.5~mHz.}
  \label{tab:EnergyRatioCenter}
  \begin{tabular}{lrrrr}
  \hline
  \multirow{2}{*}{Boundary condition}     & \multicolumn{2}{c}{3~mHz} & \multicolumn{2}{c}{6.5~mHz} \\ 
  & Energy & Ampl. & Energy & Ampl. \\ \hline
  \rowcolor{mygray}
    (Reference)          & 22.5\% & 46.7\% & \ 2.3\%& 15.6\% \\
  (Dirichlet)            & 22.5\% & 46.6\% & 23.0\%& 47.0\%   \\ 
  \eqref{eq:AtmoNL}      & 22.5\% & 46.7\% & \ 2.3\%& 15.6\%   \\
  \eqref{eq:AtmoHF1}     & 22.7\% & 46.8\% & \ 2.9\%& 15.5\%   \\
  \midrule
  \end{tabular}
  \normalsize
\end{table}

\subsection{Approximate atmospheric radiation boundary conditions}
\label{sect:val}
We now consider the case where $\KH \neq 0$ by placing the source on the rotation axis at the photosphere. In this case, the exact form of the RBCs cannot be computed in the 2D case and thus need development, as discussed in Section~\ref{sec:BCs}. Here, we compare the high-frequency and the small angle developments of \eqref{eq:AtmoNL} with the 1D setup. 
Specifically, we are interested in the error induced by considering the different approximations of \eqref{eq:AtmoNL}. 

\begin{figure*}
\centering
\includegraphics[width=0.9\linewidth]{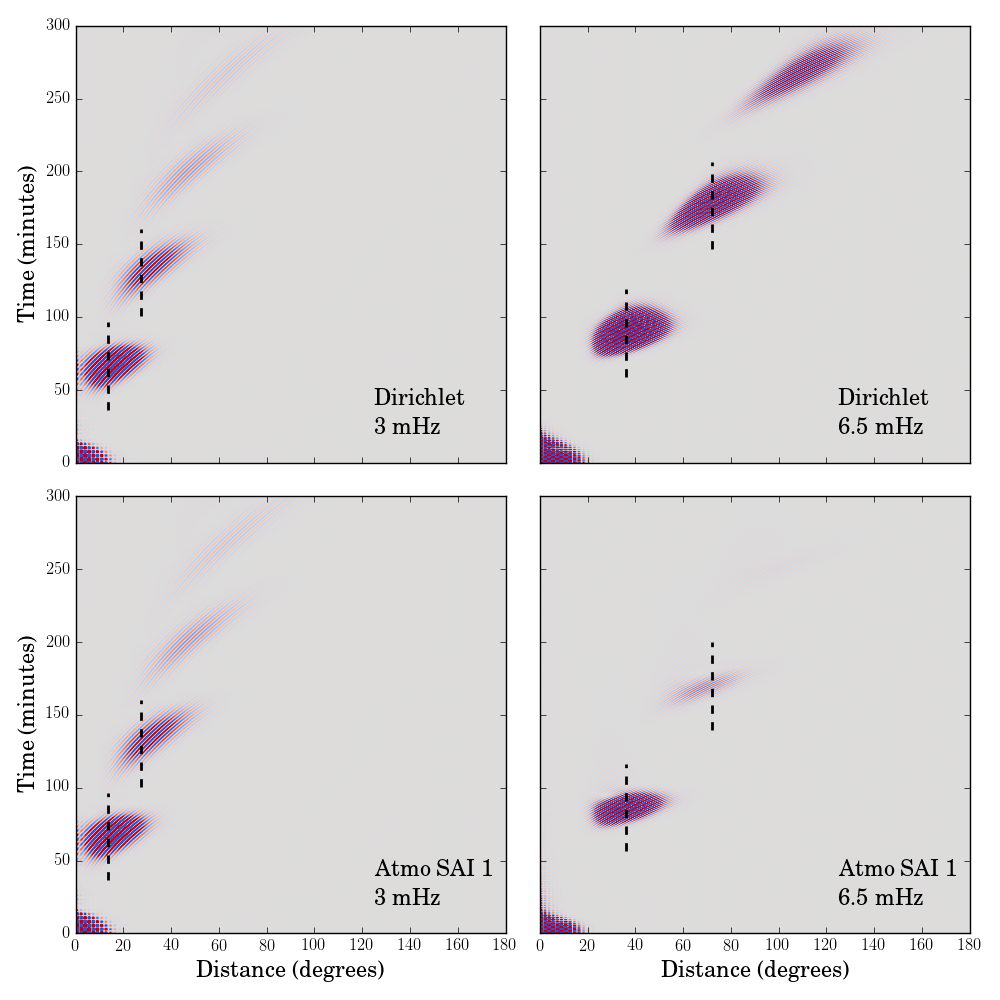}

\includegraphics[width=0.9\linewidth]{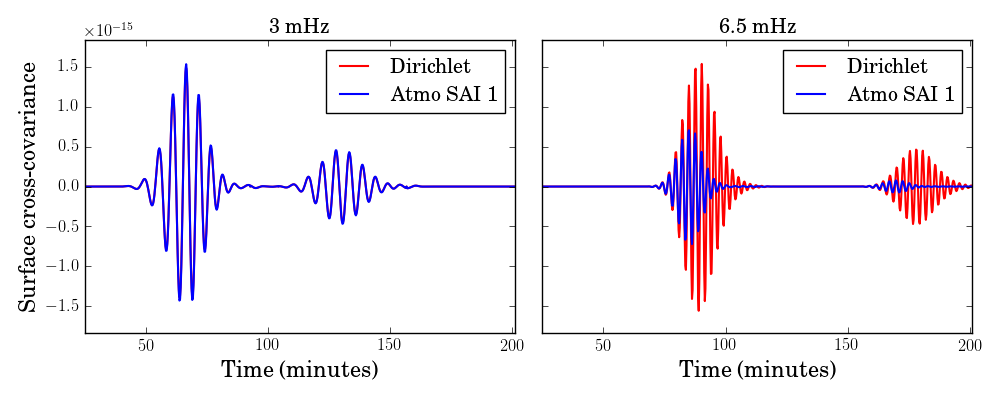}

\caption{
Time-distance diagrams $\overline{C}(\Delta, t)$ using a (Dirichlet) and \eqref{eq:AtmoHAI1} BCs by selecting the frequencies around 3~mHz (left) and 6.5~mHz (right). The cross-covariance is filtered using a phase speed filter centered at 125.2~km/s  for 3~mHz and 250.4~km/s for 6.5~mHz and a width of 12.3~km/s. The dashed black lines indicate the location of the cuts shown at the bottom in order to emphasize the amplitude ratios between the first and second skip. }
\label{fig:tdDirichlet}
\end{figure*}

Due to the fact that the source is so close to the computational boundary, it is difficult to distinguish between outgoing and ingoing waves. To address this, a phase-speed filter is applied to the power spectrum in order to select skip-distances, and so, angles of incidence. \referee{For the cross-covariance at 3 mHz the phase-speed filter is centered around 125.2~km/s. For the cross-covariance at 6.5 mHz the phase-speed filter is centered around 250.4~km/s.} A time-distance diagram for this two cases is shown in Figure~\ref{fig:tdDirichlet} for (Dirichlet) and \eqref{eq:AtmoHAI1} boundary conditions. Figure~\ref{fig:tdDirichlet} again demonstrates that the behavior of the low frequency waves is similar for both boundary conditions, due to the reflection from the density. At high frequencies the waves with Dirichlet BC show only a small diminishing in amplitude with successive skips. However, for the \eqref{eq:AtmoHAI1} BC the waves diminish rapidly with successive skips, suggesting that the high frequency waves are absorbed by the BC, as desired. Note that the amplitude of the first skip is higher for the Dirichlet BC as part of the wave packet (near the source) is reflected back due to the boundary condition. 
\begin{table}
  \small
  \centering
  \caption{  Ratios between first and second skip energies ($\Ecal_2 / \Ecal_1$) and amplitudes ($A_2 / A_1$) at the surface as illustrated in Fig.~\ref{fig:tdDirichlet}, for selected ranges of frequencies around 3~mHz and 6.5~mHz.}
  \label{tab:EnergyRatioPolar}
  \begin{tabular}{rrrrr}
  \hline
  \multirow{2}{*}{Boundary condition}    & \multicolumn{2}{c}{3~mHz} & \multicolumn{2}{c}{6.5~mHz} \\
  & Energy & Ampl. & Energy & Ampl. \\ \hline
  \rowcolor{mygray}
  (Reference)           & 11.2\% & 30.5\% & \ 1.2\%& 10.7\% \\
  (Dirichlet)           & 11.3\% & 30.6\% & 10.7\% & 30.7\%\\  
  \eqref{eq:AtmoNL}     & 11.2\% & 30.5\% & \ 1.2\%& 10.7\% \\
  \eqref{eq:AtmoHF1}    & 11.2\% & 30.6\% & \ 1.6\%& 11.9\% \\
  \eqref{eq:AtmoHAI1}   & 11.2\% & 30.5\% & \ 1.2\%& 10.7\% \\
  \eqref{eq:AtmoHAI0}   & 11.2\% & 30.5\% & \ 1.2\%& 10.7\%\\  
  \midrule
  \end{tabular}
  \normalsize
\end{table}

Table~\ref{tab:EnergyRatioPolar} details the energy and amplitude ratios between the first and second skip waves. At 3~mHz the energy of the second skip relative to the first is approximately 11\%, while the amplitude is 30\%. The difference between these skips at 3~mHz is due to the damping. However, at 6.5~mHz the RBCs show a smaller ratio between the first and second skips at 1.1\%  for the energy and 10.6\% for the amplitude. In the case of (Dirichlet), a similar ratio is reflected to that at 3~mHz. If we compare the different approximations of the BC, one can see that the approximation of a small angle of incidence gives a more accurate description of the atmosphere compared to the high-frequency approximation. The amplitude of the second peak is about $10\%$ of the first skip which is in agreement with the observations by \citet{JEF97}. 

\begin{figure}
\centering
\includegraphics[width=\linewidth]{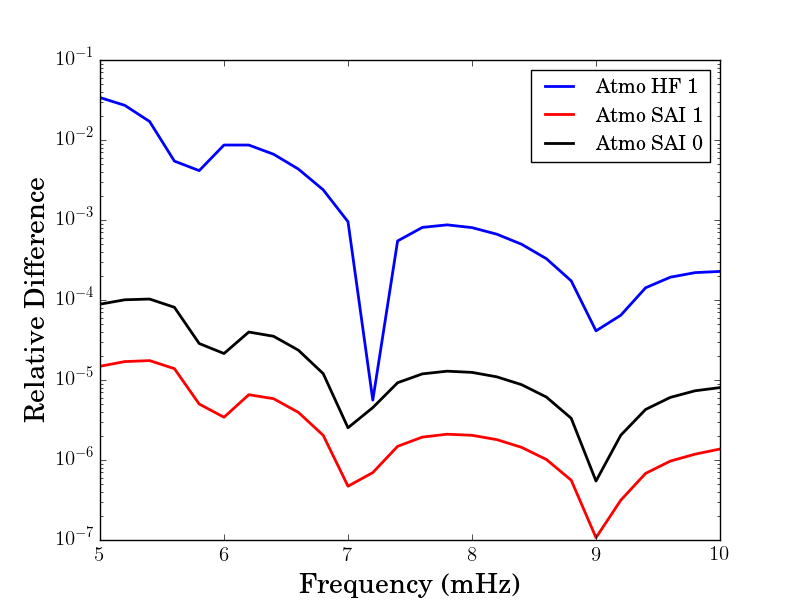}
\caption{
\referee{Relative difference between the reflection coefficients as a function of frequency. The different boundary conditions are compared with respect to the (Reference) BC. A phase speed filter centered at $250.4$~km/s is used in all cases.}
}
\label{fig:convEnergy}
\end{figure}

\referee{Since the \eqref{eq:AtmoHF1} BC is valid for very high frequencies, the accuracy of this BC as a function of frequency must be examined. The error in the reflection coefficient is shown in Fig.~\ref{fig:convEnergy}. As expected, the accuracy increases exponentially with frequency but the boundary conditions derived under the small angle of incidence assumption are still more accurate.}

\begin{figure}
\centering
\includegraphics[width=\linewidth]{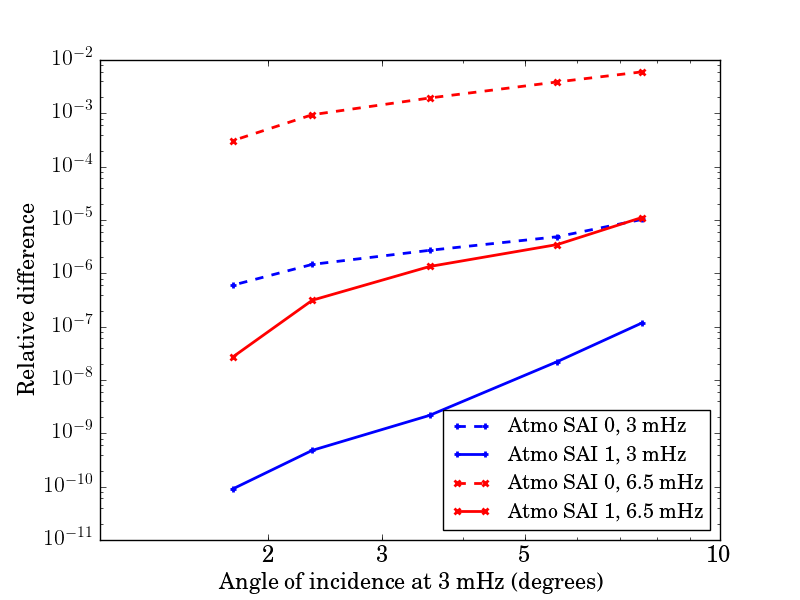}
\caption{
Relative difference between the reflection coefficients as reported in table~\ref{tab:EnergyRatioPolar} for phase speed filters ranging from $62.6$~km/s to $250.4$~km/s, i.e. angles of incidence with the normal to the surface from $1.76^\circ$ to $7.57^\circ$. The~\eqref{eq:AtmoNL} RBC energy ratio is used as reference. 
}
\label{fig:diffAngle}
\end{figure}

In order to study the impact of the angle of incidence of the wave, Fig.~\ref{fig:diffAngle} shows the difference between the \eqref{eq:AtmoHAI1} and~\eqref{eq:AtmoHAI0} RBCs with the \eqref{eq:AtmoNL} as reference, for a range of phase-speeds. We compute the angle of incidence corresponding to a specific phase-speed using ray theory for a frequency of 3~mHz. As expected from the expansions described in Sec.~\ref{sec:BC2D}, the difference decreases exponentially with the angle of incidence. However, the differences are very small which explain the tiny differences seen in the reflection coefficients in Table~\ref{tab:EnergyRatioPolar}.

\section{Importance of the atmospheric model} \label{sec:resultsHF}

So far, we have considered an atmospheric extension where the density is exponentially decaying and the sound speed is constant. However, in reality the temperature profile increases by two orders of magnitude in the transition region between the chromosphere and the corona. This increase in temperature results in an increase of sound speed and acoustic cut-off frequency by a factor of ten.
\cite{JEF97} studied time-distance diagrams at high frequencies using South Pole data and observed a double-ridge structure; they argued that it was due to the reflection of waves in this region. In this section, we study the time-distance diagram in a case when the atmosphere is extended using the VAL-C atmospheric model (see Fig.~\ref{fig:rhoc}).

\begin{figure*}
\includegraphics[width=0.95\linewidth]{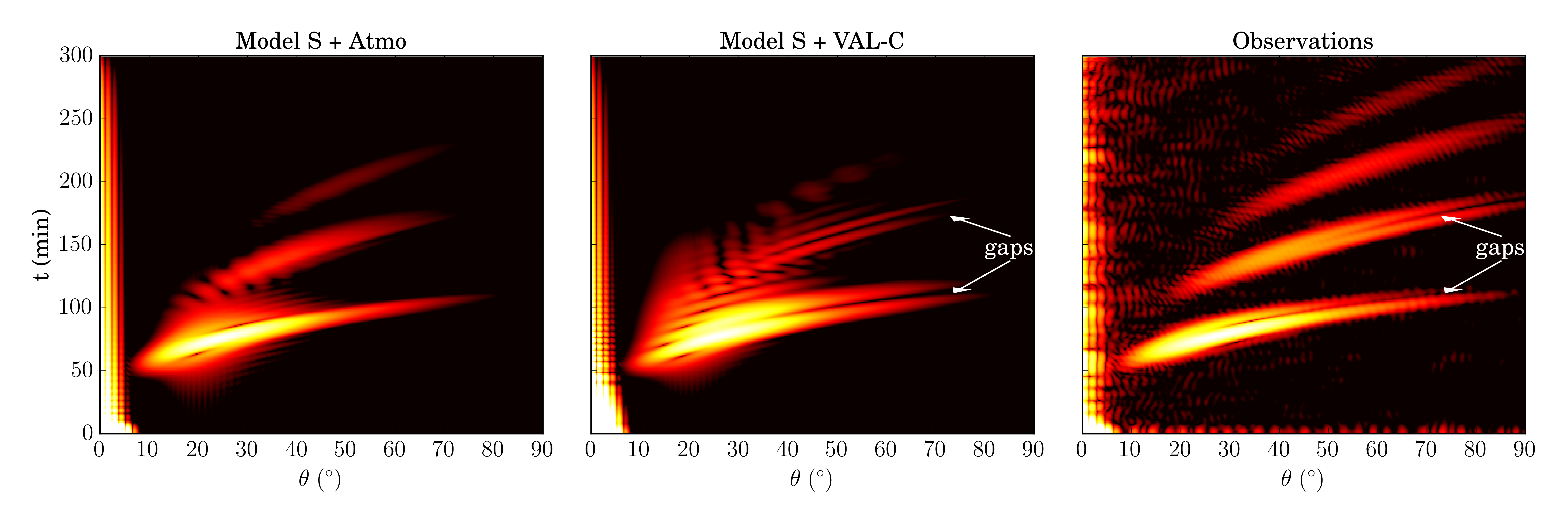}
\caption{Time-distance diagram for high-frequency waves (isolated by a Gaussian filter centered at 6.75~mHz and of FWHM 0.75~mHz) using a constant sound speed in the atmosphere (left), the VAL-C atmospheric model \citep{VER81} (center) and from MDI observations (right).} \label{fig:TD_jump}
\end{figure*}

To compare our result with the observations by \citep{JEF97}, we filter the cross-covariance with the same frequency filter and the same spatial filter. The frequency filter is a Gaussian filter centered at $6.75$~mHz with a standard deviation of $0.75$~mHz. The
spatial filter is centered at $\ell = 125$ with a standard deviation  $\sigma_\ell = 33$.  Figure~\ref{fig:TD_jump} shows the time-distance diagrams obtained with the atmospheric extension \AtmoModel \ and the VAL-C model. The models are compared to observed cross-covariance function using 72 days of MDI observations. The same frequency and spatial filters were used in all three cases.
With an upper atmosphere of constant sound speed, each skip ridge is formed of a single ridge, while a double-ridge structure is seen in the model with sound speed stratification and in the MDI observations (as observed in the South Pole data).

\citet{JEF97} suggests that the double-ridge structure is due to wave reflection at either 1 or 2 Mm above the photosphere. \citet{SEK04} noted that the double-ridge structure may be the result of an interference between waves below and above the acoustic cut-off frequency. Several improvements in our simulations would be needed to further study this question, for example by implementing a wave attenuation model that varies not only with frequency but also with height. Nevertheless,  our simulations clearly  indicate  that high-frequency waves have potential to learn about the upper solar atmosphere. They also suggest that the upper atmosphere needs to be modeled correctly in order to probe the solar interior with high-frequency waves.

\section{Conclusion}
\label{sec:conclusions}

In this paper we have applied the radiative boundary conditions developed by \citet{Barucq2017} to time-distance helioseismology, 
which were specifically developed for 
an atmosphere with a constant sound-speed and an exponentially decaying density.
The importance of such boundary conditions in computational helioseismology is in their ability to treat outgoing high-frequency waves. We found that the boundary condition derived under the assumption of small angle of incidence \ref{eq:AtmoHAI1} reproduces the most accurately the atmospheric model \AtmoModel \ and does not increase the computational burden when applied a few hundred km above the photosphere. By studying the energy of outgoing and reflected waves we have a good comparison with observations.

The radiative boundary conditions proposed here are much more elegant and less computationally expensive than extending the atmosphere by an absorbing layer (PMLs) or than transitioning to a constant medium together with a simple Sommerfeld boundary condition \citep{Gizon2017}.

In addition to the study of the boundary conditions, we have briefly examined the consequence of the sharp increase in sound speed in the transition region. Using a VAL model, we have reproduced the double-ridge feature seen in the high-frequency time-distance diagram. These results are a step in the inclusion of high-frequency waves in computational local helioseismology.

\bibliographystyle{aa}
\bibliography{aa}

\end{document}